# Large-area, low-voltage, anti-ambipolar heterojunctions from solution-processed semiconductors

*Deep Jariwala[1], Vinod K. Sangwan[1], Jung-Woo Ted Seo[1], Weichao Xu[2], Jeremy Smith[3], Chris H. Kim[2], Lincoln J. Lauhon[1], Tobin J. Marks[1,3]\*, and Mark C. Hersam[1,3]\**

[1]Department of Materials Science and Engineering, Northwestern University, Evanston, Illinois 60208, USA.

[2]Department of Electrical and Computer Engineering, University of Minnesota, Minneapolis, Minnesota 55455, USA.

[3]Department of Chemistry, Northwestern University, Evanston, Illinois 60208, USA.

\*e-mail: t-marks@northwestern.edu, m-hersam@northwestern.edu

ABSTRACT: The emergence of semiconducting materials with inert or dangling bond-free surfaces has created opportunities to form van der Waals heterostructures without the constraints of traditional epitaxial growth. For example, layered two-dimensional (2D) semiconductors have been incorporated into heterostructure devices with gate-tunable electronic and optical functionalities. However, 2D materials present processing challenges that have prevented these heterostructures from being produced with sufficient scalability and/or homogeneity to enable their incorporation into large-area integrated circuits. Here, we extend the concept of van der Waals heterojunctions to semiconducting p-type single-walled carbon nanotube (s-SWCNT) and n-type amorphous indium gallium zinc oxide (a-IGZO) thin films that can be solution-processed or sputtered with high spatial uniformity at the wafer scale. The resulting large-area, low-voltage p-n heterojunctions exhibit anti-ambipolar transfer characteristics with high on/off ratios that are well-suited for electronic, optoelectronic, and telecommunication technologies.

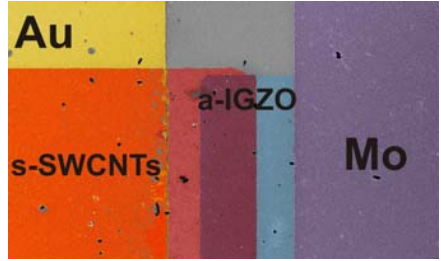

**TOC Figure**

The growing inventory of layered 2D semiconductors with diverse electronic characteristics has allowed atomically thin and dimensionally abrupt heterostructures to be realized.[1,2] Several device types including tunneling field-effect transistors (FETs),[3,4] Schottky junctions,[5] photovoltaic devices,[6,7] p-n junction diodes,[8-10] inverters,[11] and non-volatile memory cells[12] have been demonstrated with these van der Waals heterostructures. The weak interlayer bonding in these structures provides strain-free and defect-free interfaces without the constraints of epitaxy, which has accelerated the demonstration of novel device concepts and charge transport behavior. Prominent among these phenomena is anti-ambipolarity, first observed in gate-tunable, carbon-nanotube/single-layer MoS$_2$ p-n heterojunctions[9] and more recently in 2D/2D p-n heterojunctions.[10] While the examples above have employed layered 2D semiconductors using processing methods with limited scalability and uniformity, the concept of a van der Waals heterojunction can be extended to any two materials with native oxide-free surfaces[13,14] that do not covalently bond when brought in intimate contact.[15-17] Using this concept, we demonstrate here a scalable path to anti-ambipolar p-n heterojunctions by integrating two dissimilar solution-processable, air-stable materials: p-type s-SWCNTs and n-type a-IGZO. Conventional photolithography is used to fabricate large-area heterojunction arrays with high-k gate dielectrics, thereby providing low-voltage operation and device statistics at the $cm^2$ level. The anti-ambipolar behavior enables demonstration of telecommunications applications such as frequency doublers and phase shift keying circuits with reduced numbers of circuit elements compared to conventional field-effect transistor technology.

Wafer-scale heterojunction p-n diodes were fabricated using solution-processed, p-type s-SWCNTs[18] and n-type a-IGZO thin films[19-21] *via* standard photolithographic and etching techniques (Fig. 1a; see Methods and Supporting information Sections S1 and S2 for more details). The resulting device structure has an s-SWCNT FET, s-SWCNT/a-IGZO p-n heterojunction, and a-IGZO FET in series (Fig. 1b from left to right). In particular, the p-n heterojunction region between the Mo and Au electrodes is comprised of partially overlapping patterned a-IGZO and s-SWCNT films (Figs. 1 c-f).

This device architecture enables electrical characterization of the p-n heterojunction in addition to control FETs from the individual semiconductors. Figs. 2 a-b show output curves of a representative p-n heterojunction at different gate biases ($V_G$). Rectifying behavior is observed with rectification ratios exceeding $10^3$ for $V_G$ = 3 V (Supporting information Section S3). The forward current at $V_G$ = 4 V is low (~20 nA) but abruptly increases for intermediate $V_G$ values and then falls to the instrumental noise floor (~10 pA) at $V_G$ = 0 V. This behavior is also evident in the p-n heterojunction transfer plots (I-$V_G$) (Fig. 1c, green). This anti-ambipolar transfer plot shows one current maximum (on-state) in between two off-states at either extremes of the gate voltage range.[9] The voltage dependence of the anti-ambipolar plot is approximately a superposition of the transfer plots of the p-type and n-type unipolar FETs in red and blue, respectively (Fig. 2c). An anti-ambipolar response can also be produced by connecting the two unipolar FETs in series (Supporting information Section S4). However, this series geometry presents fabrication, scaling, and speed issues compared to the p-n heterojunction.

Due to screening from the ~20 nm thick a-IGZO, the junction itself is less modulated by the gate field compared to p-n heterojunctions based on 2D materials,[9, 10] thus simplifying the charge transport mechanism and facilitating the realization of reproducible and spatially homogeneous characteristics. For example, Fig. 2b shows that the reverse saturation current possesses a relatively weak gate-dependence. In Fig. 2d, three-dimensional plots of current ($I_D$) as a function of $V_G$ and forward bias voltage ($V_D$) illustrate that the charge transport is primarily a result of two semiconductors with opposite carrier types in series. For example, cross-sections along the $V_D$ axis at $V_G > 2.2$ V (point of maximum current) resemble the output plots of s-SWCNT FETs under positive bias, whereas $V_G < 2.2$ V shows a saturating behavior that correlates with the output plots of a-IGZO FETs (Supporting information Section S5). The $V_G$ dependence of the rectification ratios, band alignments, and further details on the conduction mechanism are discussed in Supporting information Section S6.

The large array of devices examined here (Fig. 1a) enables assessment of the uniformity in performance by statistical means. As observed in Fig. 3a, consistent anti-ambipolar behavior is observed among 115 devices measured on two separate chips (Fig. 3a). Two important performance metrics for an anti-ambipolar device are the position of the current maximum in terms of gate voltage ($V_{max}$) and the on/off current ratio. Since $V_{max}$ dictates the operational parameters of integrated circuits (*vide infra*), it is important for this parameter to be spatially homogenous. As seen in Fig. 3b, the histogram of $V_{max}$ peaks at 2.2 V with a tight distribution (the standard deviation is approximately 8% of the $V_G$ sweep range). Anti-ambipolar devices possess two off states, and thus Fig. 3c provides two histograms for the on/off ratio. In both cases, the $\log_{10}(I_{on}/I_{off})$ histograms have mean values > 3 with relative standard deviations of 20% and 15% for the a-IGZO and s-SWCNT sides, respectively. The high (>1000) and consistent value of these on/off ratios suggests that these devices are suitable for digital electronic applications. Details on device dimensions and variations in the junction area in the array are discussed in Supporting information Section S7.

The most important characteristic of the anti-ambipolar response curve is the presence of positive and negative transconductances on the left and right side of the current maximum, respectively. The change in the sign of the transconductance can be exploited for analog circuit applications such as frequency doubling circuits (Fig. 4a). Frequency doubling (or multiplying) circuits have broad applications ranging from analog communications to radio astronomy and THz sensing.[22] When an anti-ambipolar device is biased such that $V_{offset} < V_{max}$, the transconductance is positive with the current ($I_{DS}$) rising and the output voltage ($V_o$) across the resistor (R) increasing with the positive phase of the input signal. The case where $V_{offset} > V_{max}$ is similar except that the output signal is out of phase with the input since the input signal experiences a negative transconductance with increasing voltage. Also, when $V_G = V_{max}$, the transconductance is zero, resulting in local maxima and minima in the output signal whenever the input signal crosses $V_{max}$ in either direction. The overall effect is frequency doubling when $V_{offset}$ is set equal to $V_{max}$ (Fig. 4c). If $V_{offset}$ is moved away from $V_{max}$, the frequency doubling is incomplete, which supports our circuit operation model and enables further tunability of the output signal. Note also that the power spectral purity of the frequency doubled output signal is ~95% which exceeds the performance of graphene-based frequency doublers[22, 23] (Supporting information Section S8). Refinements such as scaling down device dimensions and local gating are likely to further enhance the anti-ambipolar frequency doubling performance (*e.g.,* enabling higher operating frequencies).

Anti-ambipolarity facilitates the realization of other analog signal processing circuits including binary phase shift keying (BPSK) circuits that are used for passband data transmission in digital communication systems.[24] These circuits map the conceptual symbols digital 0 and digital 1 into physical quantities that can be carried by alternating current (AC) signals. In this manner, BPSK is widely used for telecommunications and wireless data transmission technologies such as in the IEEE 802.11 standard,[25] commonly known as WiFi. Its main function is to modulate the carrier AC signal with no phase shift for digital 0 transmission and with a 180º phase shift for digital 1 transmission. The input is typically a sine wave superimposed on a modulating square wave signal that possesses the desired data pattern. In the anti-ambipolar BPSK circuit (Fig. 5a-b; Supporting information Section S8), the output undergoes a phase shift at every edge of the square wave when the input $V_{offset}$ is aligned with $V_{max}$. In contrast to conventional Si integrated circuit technology that achieves this circuit function using a Gilbert cell consisting of at least 7 FETs,[26, 27] the anti-ambipolar implementation requires only one p-n heterojunction in series with one resistor. By changing $V_{offset}$ and the input amplitude, another keying operation, namely binary frequency shift keying (BFSK), is demonstrated (Fig. 5c). BFSK achieves frequency doubling of the output AC signal in response to the input square wave, and is a special case of frequency modulation with applications in microwave radio and satellite transmission systems.[24] Again, the anti-ambipolar implementation requires considerably fewer circuit elements compared to conventional Si technology, thus simplifying circuit design and implementation.

The present s-SWCNT/a-IGZO p-n heterojunction demonstrates that van der Waals heterostructures are not limited to 2D semiconductors, which considerably broadens the potential of this device concept. For example, the solution-processability and ambient-stability of s-SWCNTs and a-IGZO allow reproducible anti-ambipolar devices to be achieved over large areas on arbitrary substrates using well-established manufacturing methods. In this manner, a suite of telecommunications circuits have been implemented and found to possess improved simplicity compared to established Si technology. Furthermore, this p-n heterojunction device geometry allows engineering of the anti-ambipolar transfer curve by appropriate choice of the constituent semiconductors and their respective threshold voltages, thus presenting additional opportunities for customization of the anti-ambipolar response for other circuits and systems. Future work will focus on optimizing the device geometry to minimize its lateral footprint and fringe capacitance, which will enable improvements in integration density and operating speed.

**METHODS:**

**Materials synthesis and deposition:** a-IGZO films were grown by spin-coating a combustion precursor solution and annealing on a hot plate at 300 °C for 10 min.[19] The precursor solution consisted of In, Ga, and Zn nitrates dissolved in 2-methoxyethanol (0.05 mol L$^{-1}$) with the addition of acetylacetone as a fuel and NH$_4$OH to improve acetylacetone coordination to the metal. The In:Ga:Zn ratio of 72.5:7.5:20 was chosen to optimize transistor performance.[28] The total required film thickness (~20 nm) was achieved by four repeated spin-coating/annealing steps. The s-SWCNTs were sorted using density gradient ultracentrifugation (DGU). Sorted s-SWCNT thin films were prepared by vacuum filtration followed by thorough cleaning with DI water. SWCNTs films were then transferred from a cellulose membrane onto device substrates using an acetone bath transfer technique.[29]

**Device fabrication and measurements:** Hafnia (15 nm) was deposited on a degenerately doped silicon wafer using atomic layer deposition followed by solution deposition of a-IGZO. Four steps of photolithography were used to: (1) Define Mo (100 nm) electrodes; Define a-IGZO channels using oxalic acid (10 % in water); (3) Define Ti/Au (2 nm/50 nm) electrodes; (4) Define s-SWCNT channels *via* reactive ion etching (RIE) in $O_2$. After the final step of s-SWCNT RIE etching, the devices were immersed in N-methyl-2-pyrrolidone at 80 °C for 40 min to further remove photoresist and other photolithography residues. All electrical measurements were performed under ambient conditions in the dark using source-meter (Keithley 2400), waveform generator (Agilent 33500B), and oscilloscope (Agilent 54624A) instrumentation. The gate voltage was limited to 4 V on the positive side to avoid irreversible breakdown of the hafnia dielectric.

**Structural characterization of devices:** All atomic force microscopy (AFM) images were acquired in tapping mode using a Bruker Dimension ICON system. Scanning electron microscopy (SEM) images were acquired with Hitachi SU8030 system at 2 kV using the secondary electron detector.

**ASSOCIATED CONTENT**
**Supporting Information**:
Additional details on material synthesis and electrical characterization accompanies this paper and is available free of charge via the Internet at http://pubs.acs.org.

**AUTHOR INFORMATION**
**Corresponding authors:**
*E-mail: t-marks@northwestern.edu, m-hersam@northwestern.edu

**NOTES:**

**Competing financial interests**: The authors declare no competing financial interests.

**ACKNOWLEDGEMENTS**

This research was supported by the Materials Research Science and Engineering Center (MRSEC) of Northwestern University (NSF DMR-1121262), the Office of Naval Research (ONR MURI N00014-11-1-0690), and the National Institute of Standards and Technology (NIST CHiMaD 70NANB14H012). D. J. acknowledges additional support from an SPIE education scholarship, and J-W.T.S. acknowledges support from the National Science Foundation (DMR-1006391). This work made use of the Northwestern University Micro/Nano Fabrication Facility (NUFAB) and the Northwestern University Atomic and Nanoscale Characterization Experimental Center (NUANCE), which has received support from the MRSEC (NSF DMR-1121262), Nanoscale Science and Engineering Center (NSF EEC-0118025/003), State of Illinois, and Northwestern University. The authors thank Ken Everaerts, Jianting He, and Heather Arnold for helpful advice on device fabrication and material deposition processes.

**FIGURES:**

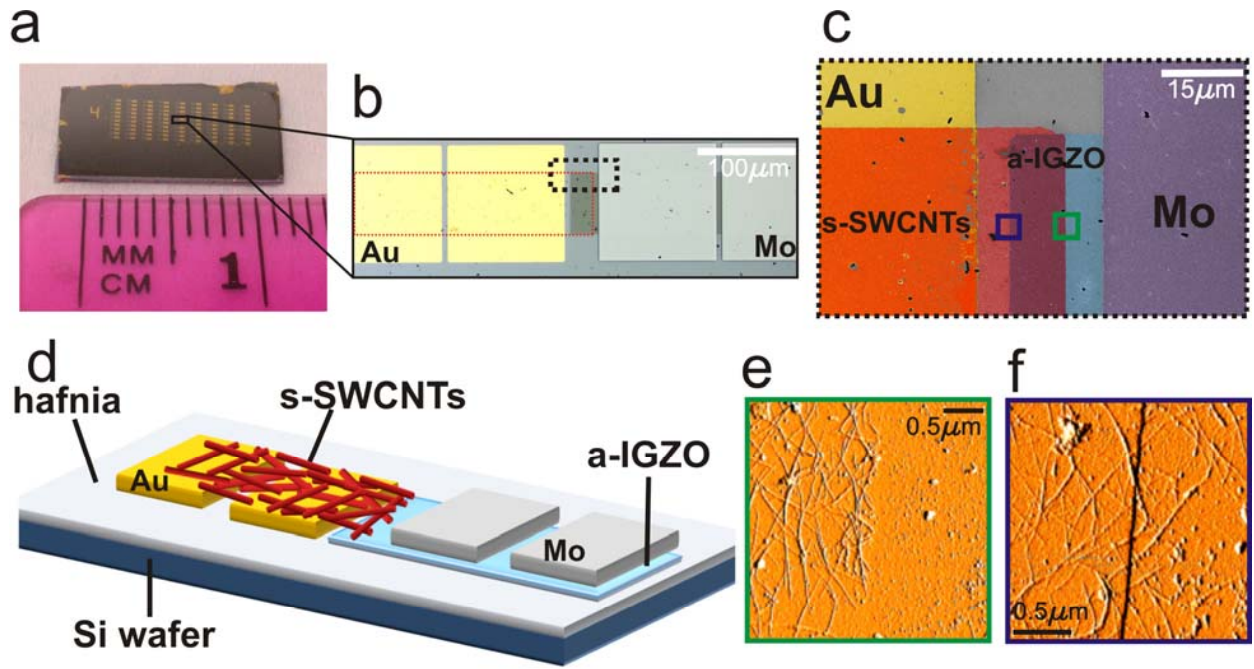

**Figure 1. Structure of the s-SWCNT/a-IGZO p-n heterojunctions. a,** Optical image of a fully fabricated p-n heterojunction device array. **b,** Expanded optical micrograph of a representative individual device in the array. (left to right) p-type s-SWCNT FET, s-SWCNT/a-IGZO p-n heterojunction, and n-type a-IGZO FET. The dashed red outline indicates the patterned s-SWCNT thin film, while the dark rectangular region in the channel is the patterned a-IGZO thin film. **c,** False-color SEM image of the dashed black outline in **b.** The different layers are appropriately colored and labeled. The Au electrode (drain) is biased while the Mo electrode (source) is grounded. **d,** Schematic diagram of an individual unit of the array shown in **a** with appropriately labeled layers. **e-f,** Atomic force micrographs (amplitude error) of the regions denoted by the green and purple squares in **c**. The patterned boundary of the s-SWCNT thin film is visible in **e** while the black line in **f** represents the patterned edge of the a-IGZO film.

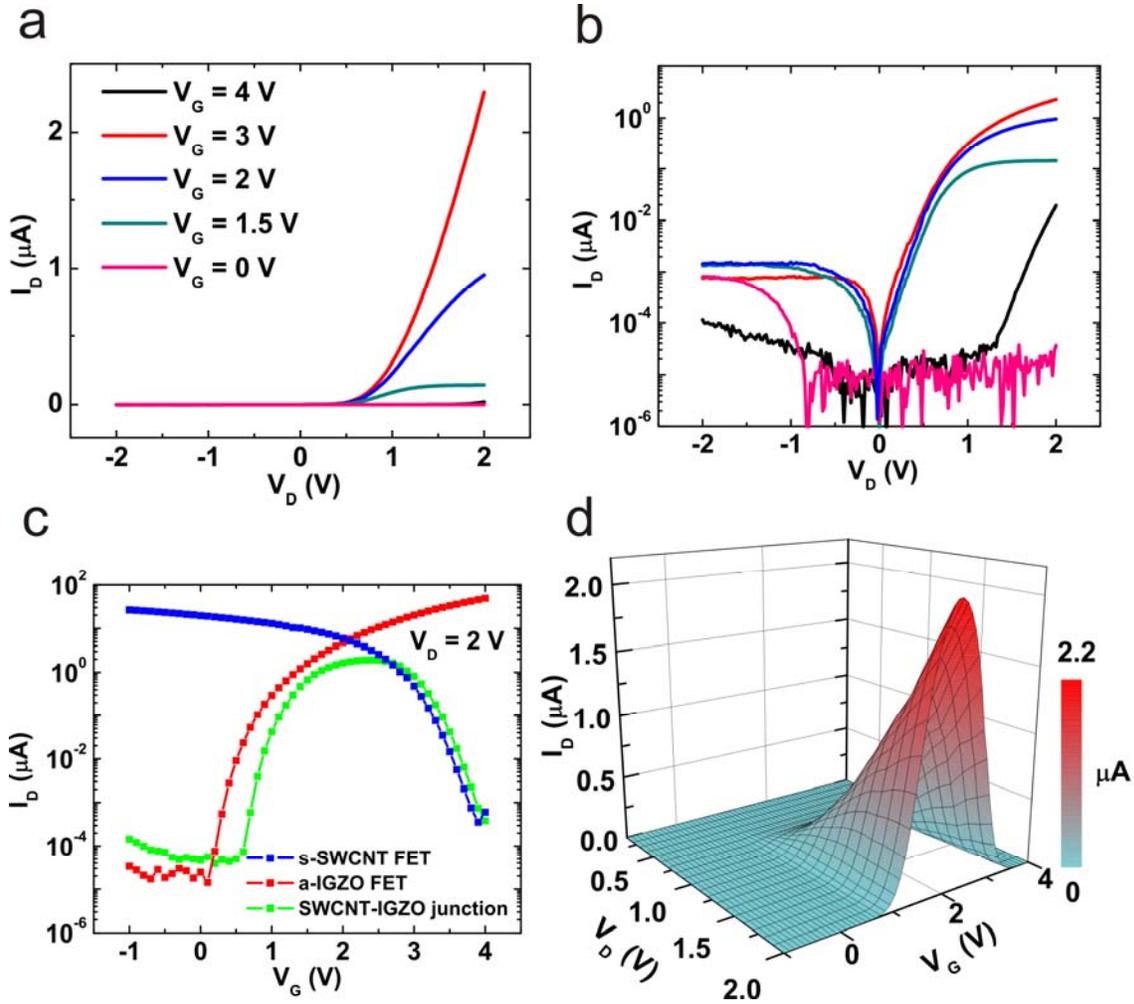

**Figure 2. Electrical properties of the s-SWCNT/a-IGZO anti-ambipolar p-n heterojunctions. a-b,** Output characteristics of a representative device at different gate voltages on linear (**a**) and semi-log (**b**) y-axis. The device is in a nearly insulating state at $V_G$ = 4 V and 0 V, while it shows a highly rectifying state at the intermediate gate voltages. The weak gate modulation of the reverse saturation current magnitude can be seen in **b**. The plot colors in **a** and **b** represent the same gate voltage values as indicated in the legend of **a**. **c,** Semi-log transfer characteristics of a p-type s-SWCNT FET (blue), n-type a-IGZO FET (red), and s-SWCNT/a-IGZO p-n heterojunction (green). **d,** Three-dimensional representation of the anti-ambipolar transfer characteristics at varying drain biases. The grid lines running along the $V_D$ axis represent the forward output characteristics at the indicated gate voltage ($V_G$).

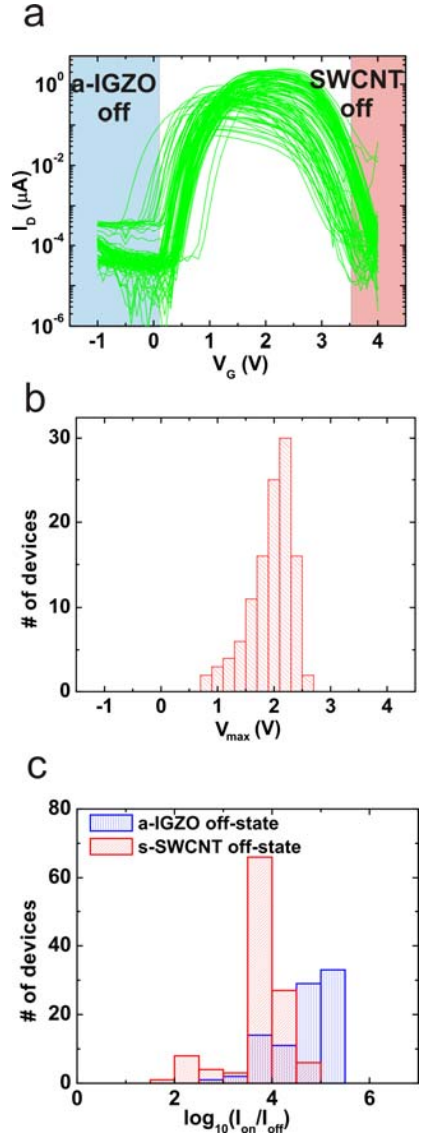

**Figure 3. Performance metric statistics of the s-SWCNT/a-IGZO anti-ambipolar p-n heterojunctions. a,** Anti-ambipolar transfer characteristics of 115 separate devices. Each curve has two off states and a current maximum (on-state) between them. **b,** Histogram of the gate voltages corresponding to the current maxima (mean = 1.89 V; standard deviation = 0.38 V). **c,** Histograms of the on/off current ratios. The red and blue histograms correspond to the ratios derived using off-currents for the s-SWCNT (mean = 3.76; standard deviation = 0.94) and a-IGZO (mean = 4.55; standard deviation = 0.94), respectively. The average on/off ratio on the a-IGZO off side is higher due to the lower off-currents in the a-IGZO off-state as seen in **a**.

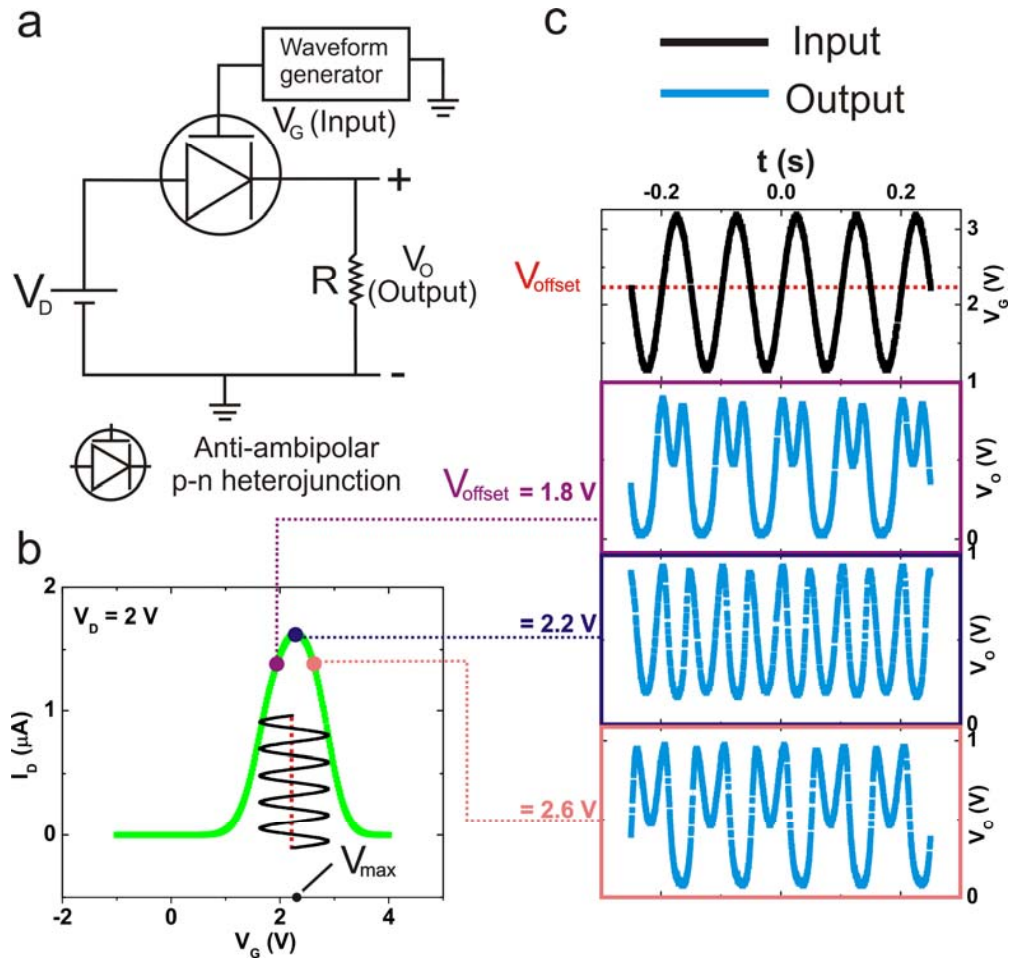

**Figure 4. Frequency doubling circuit based on s-SWCNT/a-IGZO anti-ambipolar p-n heterojunctions. a,** Schematic of the circuit employing an anti-ambipolar heterojunction for frequency doubling. The circuit uses a single anti-ambipolar heterojunction in series with a resistor across which the output voltage ($V_o$) is measured using an oscilloscope. Series resistance R = 1 MΩ. The resistance was chosen to maintain an output voltage of 1 V when the junction resistance is minimized. **b,** Representative transfer characteristic of an anti-ambipolar heterojunction. The offset voltage that is applied to the sinusoidally varying input is indicated by the differently colored circles. **c,** Input signal (black) and output signal (blue) for the three different values of the offset voltage indicated in **b.** Complete frequency doubling is observed

when $V_{offset} = V_{max}$. More complicated signal conditioning occurs when $V_{offset}$ is tuned away from $V_{max}$.

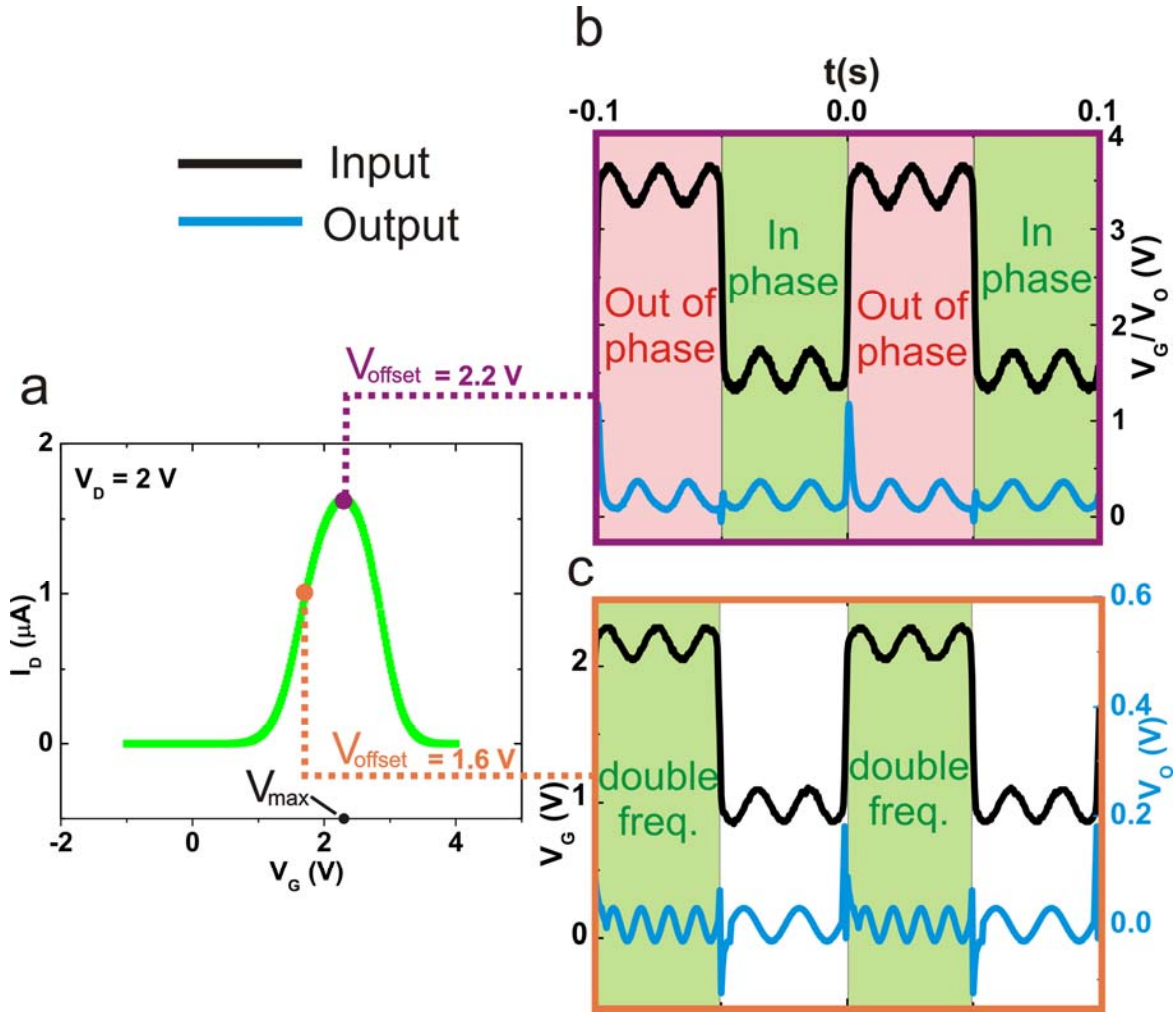

**Figure 5. Phase and frequency shift keying using s-SWCNT/a-IGZO anti-ambipolar p-n heterojunctions. a,** Representative anti-ambipolar transfer characteristic with the voltage offsets of the input signals (sine wave superimposed on modulating square waves) indicated by the differently colored circles. **b,** Binary phase shift keying (BPSK) operation using a square + sine input (black) with a voltage offset corresponding to $V_{max}$. The output sine wave (blue) shows a phase shift compared to the input sine wave for each half of the square wave modulation. **c,** Binary frequency shift keying (BFSK) operation using a square + sine input (black) with a voltage offset away from the current maximum as indicated by the red circle in **a**. The frequency of the output signal (blue) is doubled for every alternate modulation of the square wave.

# Supporting Information

# Large-area, low-voltage, anti-ambipolar heterojunctions from solution-processed semiconductors

*Deep Jariwala[1], Vinod K. Sangwan[1], Jung-Woo Ted Seo[1], Weichao Xu[2], Jeremy Smith[3], Chris H. Kim[2], Lincoln J. Lauhon[1], Tobin J. Marks[1,3]\*, and Mark C. Hersam[1,3]\**

[1]Department of Materials Science and Engineering, Northwestern University, Evanston, Illinois 60208, USA.

[2]Department of Electrical and Computer Engineering, University of Minnesota, Minneapolis, Minnesota 55455, USA.

[3]Department of Chemistry, Northwestern University, Evanston, Illinois 60208, USA.

*e-mail: t-marks@northwestern.edu, m-hersam@northwestern.edu

## S1. Material synthesis, deposition and characterization:

Arc discharge single-walled carbon nanotubes (P2, Carbon Solutions) were used for the preparation of semiconducting single-walled carbon nanotubes (s-SWCNTs). In particular, 45 mg of raw SWCNT powder was added to 6.6 mL of 1% w/v aqueous sodium cholate (SC) solution in a glass vial and then sonicated using a horn ultrasonicator with a 0.125'' diameter probe (Fisher Scientific 500 Sonic Dismembrator) for 1 hour at 20% of the maximum tip amplitude. Heating of the vial was minimized through the use of an ice/water bath. Following sonication, additional 1% w/v aqueous SC and sodium dodecyl sulfate (SDS) solutions containing 60% w/v iodixanol were added to the SWCNT dispersion to obtain a final iodixanol concentration of 32.5% w/v and surfactant ratio of 1:4 (SDS:SC). The SWCNT dispersion was subsequently centrifuged at 3000 rpm for 3 minutes to eliminate large SWCNT aggregates and carbonaceous impurities. Then, 6 mL of the SWCNT dispersion was inserted below 15 mL linear density gradient of 15-30% w/v iodixanol (1.08-1.16 g/ml) by using a syringe pump, and the remainder of the ultracentrifuge tube was filled with 0% w/v iodixanol aqueous solution. The entire gradient contained a 1:4 ratio of 1% w/v SDS:SC. The linear density gradients were then ultracentrifuged for 18 hours at 32 krpm in an SW 32 rotor (Beckman Coulter) at a temperature of 22 °C. The resulting layer of s-SWCNTs at the top of the gradient was extracted using a piston gradient fractionator (Biocomp Instruments).

The electronic purity of s-SWCNTs was estimated by measuring their optical absorbance spectra with a Cary 5000 spectrophotometer (Agilent Technologies) (Fig. S1). For the measurement, the extracted SWCNT fractions were diluted in disposable plastic cuvettes (Fisher Scientific). Reference solutions were made with 1% w/v of SDS and SC (1:4 ratio) with addition of iodixanol for the baseline measurement. The electronic purity of the s-SWCNTs was determined from their optical absorbance spectra after subtracting the π-plasmon resonance contributions and linear background with respect to energy, and then comparing the area of the second-order semiconducting peak ($S_{22}$) to the first-order metallic peak ($M_{11}$). The resulting semiconducting purity was estimated to be ~99%.

Separate a-IGZO precursor solutions were made for In, Ga and Zn using $In(NO_3)_3 \cdot 3H_2O$, $Ga(NO_3)_3 \cdot 8H_2O$, and $Zn(NO_3)_2 \cdot 6H_2O$, respectively, dissolved in anhydrous 2-methoxyethanol (0.05 mol $L^{-1}$). To these were added acetylacetone (0.03125 mol $L^{-1}$) and aqueous ammonium hydroxide (0.0425 mol $L^{-1}$) after which the solutions were stirred overnight at room temperature. Prior to spin-coating, they were combined in the correct molar ratio (In:Ga:Zn = 72.5:7.5:20), stirred for an additional 1 hour, and filtered through a 0.2 μm PTFE filter. Spin coating was carried out at 3500 rpm for 30 sec in air with < 25 % relative humidity after which the films were immediately annealed at 300 °C on a hot plate for 10 min. This process was repeated 4 times to give the overall required thickness. [1,2]

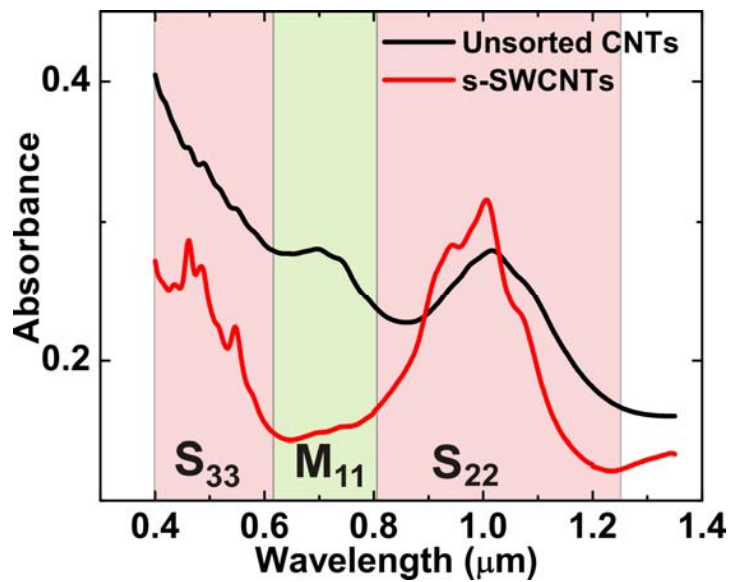

**Figure S1:** Optical absorbance spectra of large diameter s-SWCNTs (red) and unsorted carbon nanotubes (black). The $S_{22}$, $S_{33}$ (semiconducting), and $M_{11}$ (metallic) transitions are indicated by the pink and green shaded regions, respectively. The near elimination of the $M_{11}$ transitions in the red spectrum indicates high semiconducting purity of the sorted material.

**S2. Device fabrication and electrical characterization:**

All devices were fabricated on heavily doped Si <100> (n-doped, resistivity = 0.02 Ω-cm, WRS Materials) substrates. The as-received wafers were sonicated in acetone followed by isopropanol for 5 min each and immediately blow dried in nitrogen. Atomic layer deposition (ALD) was performed on the clean Si chips using a Savannah S100 ALD reactor (Cambridge Nanotech, Cambridge MA). The substrates were loaded into the chamber, which was pre-heated to 100 °C, and then the chamber temperature was increased to 200 °C for the duration of growth. The samples were exposed to sequential doses of the oxide precursor and deionized water interspersed with dry $N_2$ purge steps between each precursor dose. The precursor for $HfO_x$ films was tetrakis(dimethylamido)hafnium(IV) (TDMAHf, Aldrich, 99.99%), which was maintained at a constant temperature of 75 °C. A single ALD cycle consisted of a TDMAHf pulse for 0.25 s and a 10 s purge, followed by a H$_2$O pulse for 0.015 s and another 10 s purge, which resulted in a growth rate of ~1 Å/cycle. These conditions were used to grow ~15 nm thick HfO$_x$, which has a capacitance of 730 nF/cm$^2$.

a-IGZO was deposited on the HfO$_x$-coated Si substrates as described above in section S1. Photolithography (negative resist NR9-1000 PY, Futurex) was performed using a standard mask aligner (Suss MAB-A6) to define electrode patterns on the a-IGZO followed by sputter deposition (AJA Orion) of Mo (~100 nm) and liftoff in n-methyl-2-pyrolidone (NMP). Subsequently, another photolithography step (positive resist, S1813 Shipley Microposit) was used to define the a-IGZO patterns and etch them using a 10% oxalic acid in water solution. A third photolithography step (negative) allowed the Au electrodes to be defined for the s-SWCNT film. In particular, a thermal evaporator was used to evaporate 50 nm thick Au on top of a 2 nm thick Ti adhesion layer. The s-SWCNT films were prepared by vacuum filtration through a cellulose membrane filter, which were then stamped onto the substrates and held in an acetone bath. The density of nanotubes on the filter was controlled by the amount of solution filtered through the membrane.[3] Following a fourth photolithography step, reactive ion etching (Samco RIE-10 NR) in an oxygen plasma atmosphere (100 mW, 15 s, 20 sccm) was used to define the nanotube channels. The resist was subsequently dissolved in hot (80 °C) N-methyl-2-pyrrolidone for 2 h. No annealing step was required for the s-SWCNT films or the final device.

All measurements were carried out in ambient using a standard probe station with micromanipulators (Cascade Microtech), source meters (Keithley 2400), waveform generator (Agilent 33500B), and oscilloscope (Agilent 54624A). The schematic diagrams of the measurement setups are shown in Fig. S2 a, c, e, while the corresponding transfer plots are provided in Fig. S2 b, d, f. The hysteresis is noticeable in the transfer plot of the s-SWCNT FET (Fig. S2 b), while it is negligible for the a-IGZO FET (Fig. S2 d). The anti-ambipolar transfer plot of the junction also has noticeable hysteresis, particularly on the side where the s-SWCNTs are dominating the resistance. This asymmetric hysteresis provides further evidence of the anti-ambipolarity resulting from a series connection between s-SWCNTs and a-IGZO.

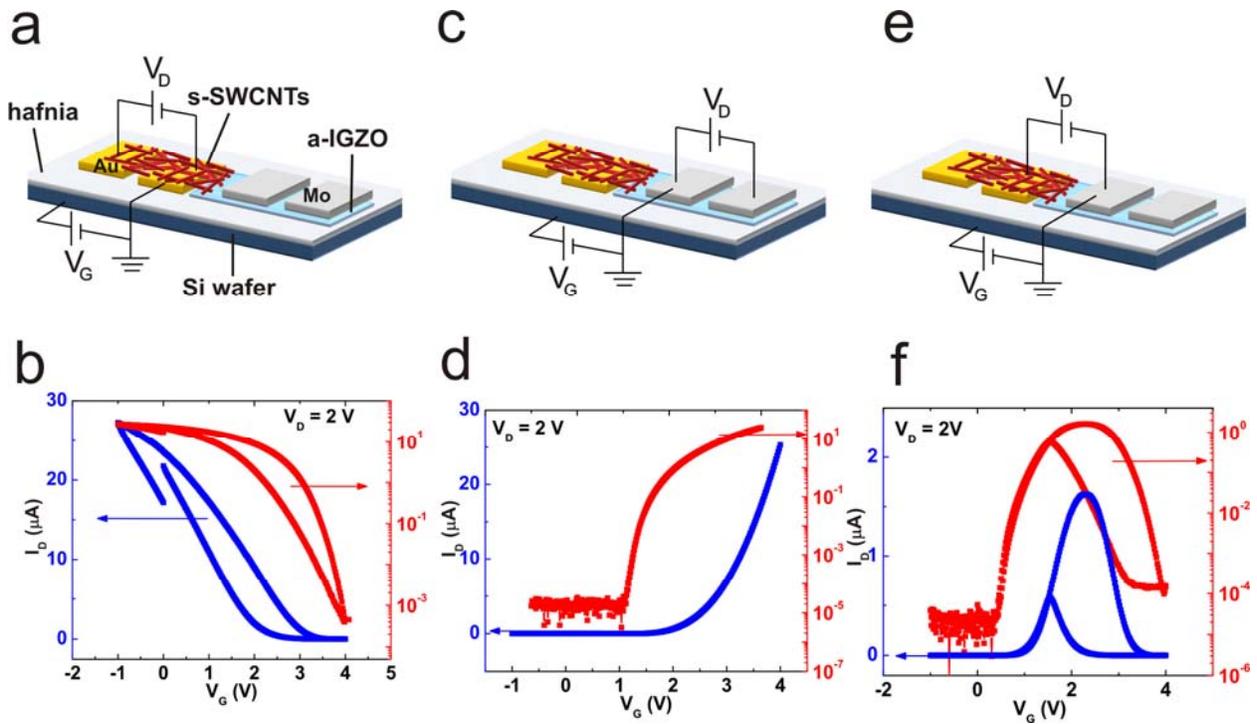

**Figure S2:** (a) Schematic and (b) linear (blue) and semi-log (red) transfer plots of a p-type s-SWCNT FET. The field-effect mobility ($\mu_{FE}$) estimated from the linear plot is ~1 cm$^2$/V.s. (c) Schematic and (d) linear (blue) and semi-log (red) transfer plots of a n-type a-IGZO FET. The field-effect mobility ($\mu_{FE}$) estimated from the linear plot is ~1 cm$^2$/V.s. (e) Schematic and (f) linear (blue) and semi-log (red) transfer plots of a s-SWCNT/a-IGZO p-n heterojunction.

## S3. Fitting output curve to the diode equation:

The output characteristics were fit to the standard Shockley diode equation. The s-SWCNT/a-IGZO p-n heterojunction diodes possess an ideality factor of 2.3 for $V_G = 3$ V. The ideality factors were > 3 for the other gate voltages. Although the ideality factor is relatively large, it is still comparable or better than the recently reported 2D/2D black phosphorus/MoS$_2$ p-n heterojunction diodes.[4] The large ideality factor can likely be attributed to traps at the s-SWCNT/a-IGZO interface.[5]

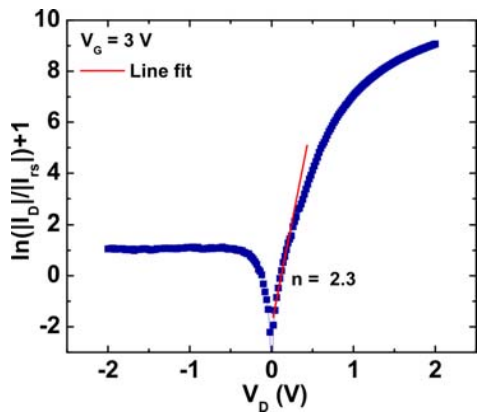

**Figure S3:** Reverse saturation current normalized semi-log output plot at $V_G = 3$ V. The linear fit at low drain bias yields an ideality factor n = 2.3.

## S4. Measuring s-SWCNT and a-IGZO FETs in series:

Our device geometry allows us to compare the junction transfer characteristics with the unipolar FETs on either side of the junction. Similarly, it allows us to connect the two unipolar FETs in series by shorting the two electrodes of the junction externally with connecting wires (Fig. S4 a). This arrangement eliminates the junction resistance (as well as the rectification property of the p-n heterojunction) while still giving an anti-ambipolar transfer behavior (Fig. S4 b-c). The current in this configuration is higher than the junction current in the linear part of the transfer curve with the maximum current being more than twice that of the junction maximum. The gate voltage at the maximum current ($V_{max}$) is nearly same (within 0.1 V) as that of the junction, which further confirms the hypothesis that the junction I-V is a result of two semiconductors in series with an additional voltage-tunable resistance at the junction interface. The series FET geometry, however, is less scalable due to the larger lateral footprint of two FETs compared to a single junction. The additional contact in the series FET geometry will also contribute additional capacitance that will compromise the ultimate speed of a fully scaled device. Furthermore, the use of different contact metals for the two FETs complicates the design of the interconnect in the series FET geometry. Consequently, for high-performance electronics, the p-n heterojunction geometry is likely to have significant advantages compared to series FETs.

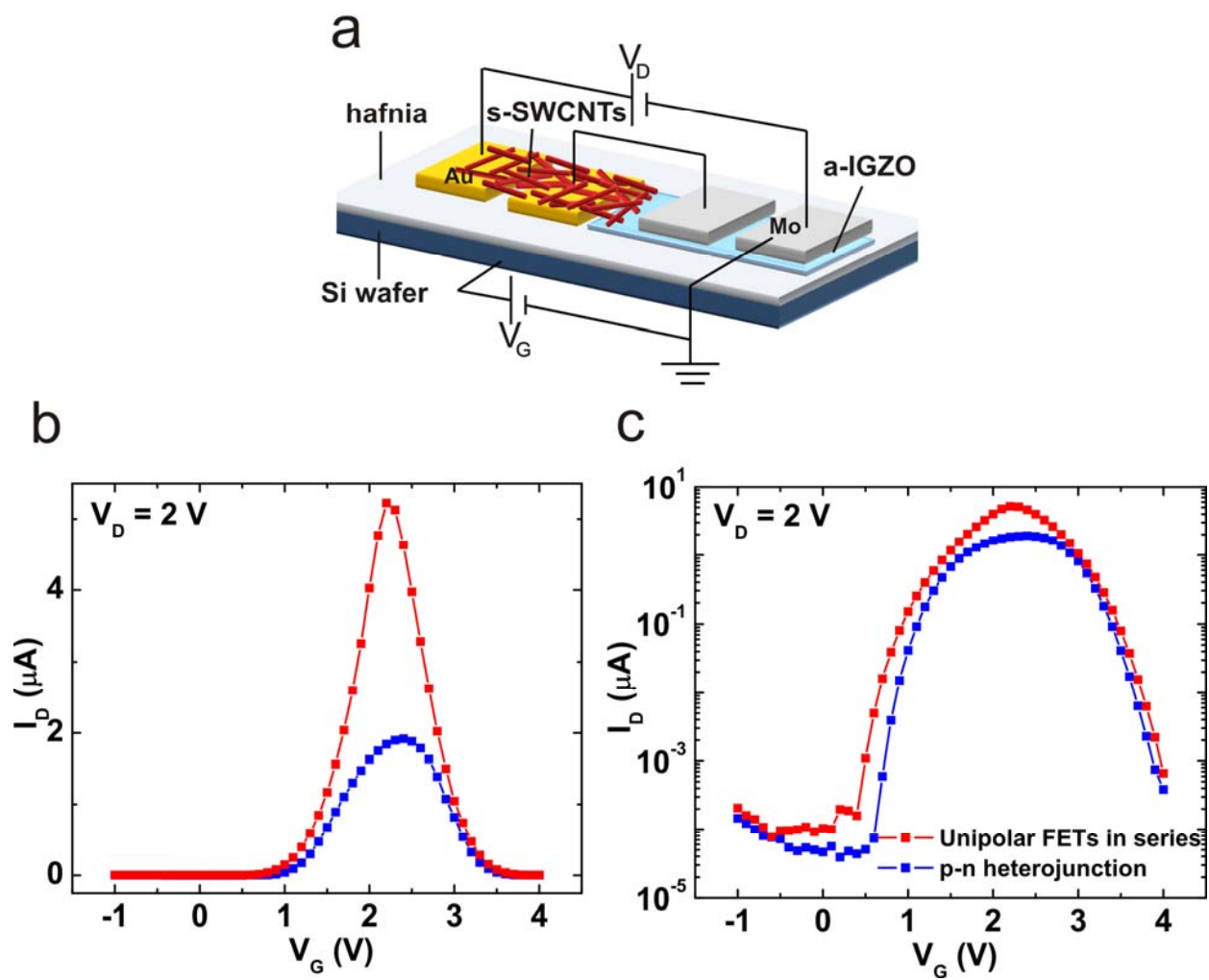

**Figure S4:** Comparison of junction transfer characteristics with unipolar FETs in series. (a) Linear and (b) semi-log transfer characteristics. The differences in the magnitude of the peak current can be attributed to the junction resistance.

## S5. $I_D$-$V_D$ characteristics of unipolar FETs:

The output characteristics of the s-SWCNT and a-IGZO unipolar FETs (Fig. S5) at positive $V_D$ values resemble those of the junction in forward bias (Fig. 2d). At $V_G >$ $V_{max}$ (2.2

V), the junction current correlates with the linearly rising current of the s-SWCNT FET at positive $V_D$ (Fig. S5 a). At $V_G < V_{max}$ (2.2 V), the junction current shows saturation behavior similar to the a-IGZO FET output plot in Fig. S5 b. Thus, the heterojunction current is dominated by the most resistive element (i.e., the one carrying the lower current) at any given voltage.

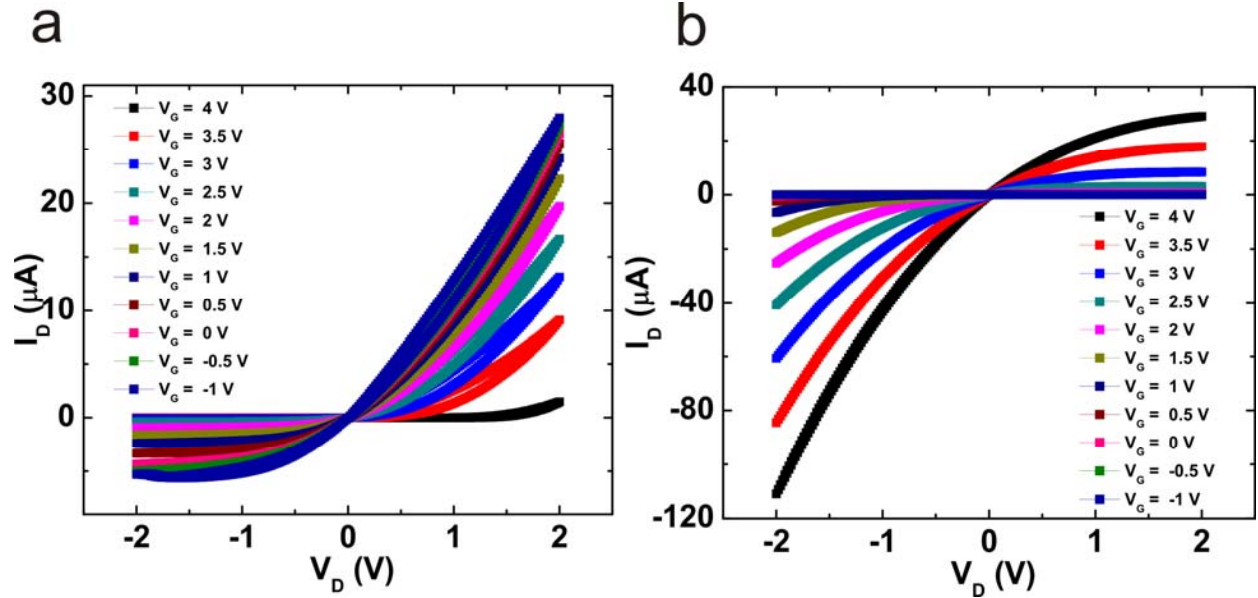

**Figure S5:** (a) Linear output plot of the network s-SWCNT FET (p-type) at varying gate biases. (b) Linear output plot of the a-IGZO FET (n-type) at different gate biases.

**S6. Gate voltage dependence of rectification ratios, band alignments, and conduction mechanisms:**

Due to the relatively large thickness (~20 nm) and high dielectric constant[6] of the a-IGZO film, the band alignment at the junction is not subject to strong modulation by the gate electric field. This expectation is supported by the weak gate modulation of reverse saturation currents as shown in Figure 2b of the manuscript. The gate dependence of the rectification ratios (i.e., the ratios of forward to reverse drain currents at the same bias magnitude) provides further insight into the conduction mechanism. Since the reverse bias current is not strongly modulated by the gate, the rectification ratios closely track the forward bias currents (i.e., the anti-ambipolar behavior). As seen in Figure S6 below, the rectification ratios are reduced at either extreme of the gate voltage with a maximum in the middle. From the perspective of gate-modulated doping, high positive $V_G$ causes the a-IGZO to be heavily n-doped while the s-SWCNTs are depleted to near intrinsic levels. For intermediate $V_G$ values, both the a-IGZO and s-SWCNTs have finite doping levels and hence substantial current flows at forward bias while current rectification occurs at reverse bias due to the built-in potential at the junction. Finally, at $V_G$ near 0 V and below, the a-IGZO is completely depleted. Due to the large band gap of a-IGZO, the depleted state of a-IGZO is highly insulating and thus minimizes current flow through the device.

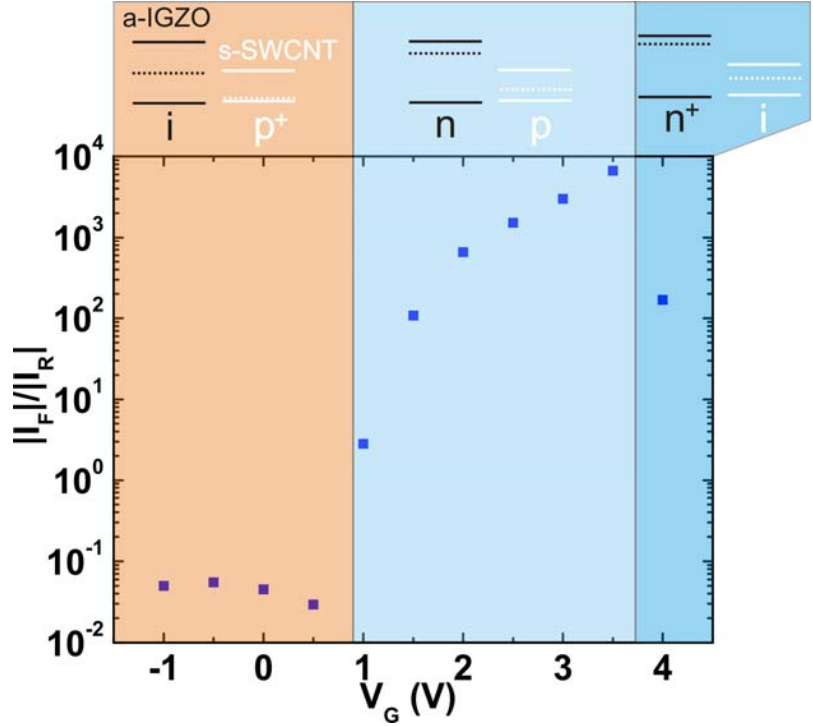

**Figure S6:** Variation of rectification ratios (at $|V_D| = 2$ V) as a function of gate voltage ($V_G$) along with corresponding band alignments in the a-IGZO and s-SWCNTs in that range.

**S7: Device dimensions and performance variability:**

A basic block diagram of a s-SWCNT/a-IGZO junction is shown below in Figure S7. $L_j$ and $W_j$ represent the length and width of the junction while $L_s^{CNT}$ and $L_s^{IGZO}$ represent the series resistance lengths of s-SWCNT and a-IGZO films, respectively, which are fixed at 5 μm in the device array. $L_j$ increases from 8 μm to 48 μm with increments of 5 μm over 9 columns. $W_j$ increases from 50 μm to 140 μm with increments of 10 μm over 10 rows. The junction area ($A_j = L_j * W_j$) thus varies from 400 μm² to 6720 μm².

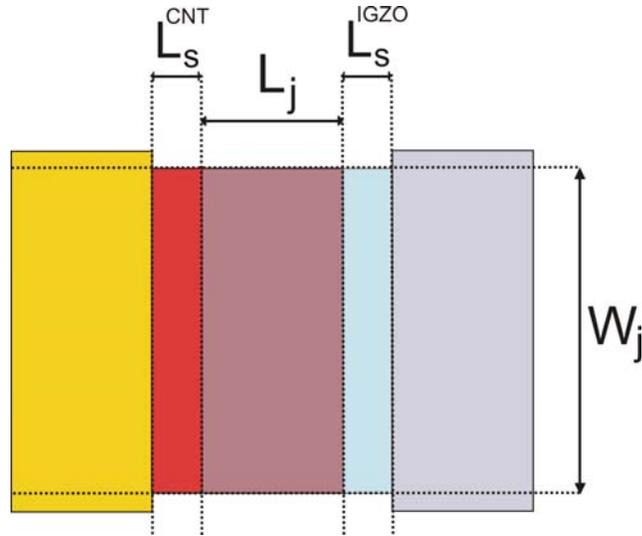

**Figure S7:** Schematic block diagram of a s-SWCNT/a-IGZO junction. The yellow and grey regions represent Au and Mo electrodes, respectively. The red and light blue regions represent s-SWCNTs and a-IGZO patterns while the overlapping area represents the junction. The schematic is similar to Figure 1 c.

Variability in device performance mainly results from variations in the structure of constituent semiconductor films. The a-IGZO, which is deposited via spin coating, is subject to thickness variations due to edge effects during spinning as well as variations in the wettability of the surface. The edge effect leads to a thicker a-IGZO film at the edges of the wafer/chip, thereby leading to higher off currents and lower on/off current ratios as evident from the histogram in Figure 3c. The nanotube film on the other hand is deposited via vacuum filtration and acetone bath transfer as described in Section S2. The nanotubes are prone to bundling and agglomeration during filtration and transfer, which leads to inhomogeneities in tube density. Small variations in tube densities leads to variations in the electrical properties of the films.[3, 7]

**S8. Analysis of frequency doubling and binary phase shift keying circuits:**

The power spectrum of the input and output signals of the frequency doubling circuit (Fig. 4c) were obtained by taking their Fourier transforms. The area under the expected output frequency was then evaluated as a percentage of the total output power. The desired output frequency of 20 Hz possessed 94.5% of the total output power as compared to ~90% for graphene frequency multipliers. It is worth noting that the s-SWCNT/a-IGZO anti-ambipolar heterojunctions are back gated and have a large device size, which contributes significantly to the parasitic capacitance. Scaling down the device dimensions and introducing local gates (as has been done in the case of graphene[8, 9]) is expected to further improve the performance of s-SWCNT/a-IGZO heterojunctions as frequency doublers.

The phase shift in a binary phase shift keying (BPSK) circuit can be defined with respect to the input signal or within the output wave itself corresponding to the phase difference before and after the square wave modulation in the input signal. The phase shift is estimated by measuring distances between peaks before and at the modulation, taking their ratios and multiplying them by 360º or by comparing the change in position of crest and troughs. The phase difference between the input and output signal is exactly 180º after the square modulation, although the phase shift (difference) in the output signal before and after the modulation is ~153º. This deviation from 180º is mainly due to the large capacitive coupling and slow response in the constituent semiconductors, which can be substantially reduced with reduced device dimensions and local gating. It should be noted, however, that the deviation in this case has negligible impact on the function of the BPSK circuit.[10]